\documentclass[10pt]{article}
\usepackage[dvips]{epsfig}
\usepackage[T1]{fontenc}
\usepackage[latin1]{inputenc}
\usepackage{graphicx}
\usepackage[english]{babel}
\usepackage{amsmath}
\usepackage{amssymb}
\usepackage{amsfonts}
\usepackage[T1]{fontenc}
\usepackage{color}
\usepackage{babel}
\usepackage{verbatim}
\usepackage[unicode=true,pdfusetitle,bookmarks=true,bookmarksnumbered=false,bookmarksopen=false,
 breaklinks=false,pdfborder={0 0 1},backref=false,colorlinks=true]{hyperref}
\hypersetup{linkcolor=blue,citecolor=blue}
\makeatletter
\usepackage[dvips]{epsfig}
\usepackage[T1]{fontenc}
\usepackage{color}
\usepackage{pdflscape}
\begin{document}

\title{\bf    Cosmological Singularities in Conformal Weyl Gravity }
\author{ Yaghoub Heydarzade\thanks{%
email: yheydarzade@bilkent.edu.tr}   
\\{\small Department of Mathematics, Faculty of Sciences, Bilkent University, 06800 Ankara, Turkey}}
\date{\today} 
\maketitle
\begin{abstract}
In this work, we study the issue of the past and future cosmological singularities
in the context of the fourth-order conformal Weyl gravity.
In particular, we investigate the emergent universe scenario proposed by
Ellis {\it et al}, and find the stability conditions of the corresponding Einstein static
state using the fixed point
approach.
We show that depending on the values of the parameters of the conformal Weyl
gravity theory, there are possibilities for having initially stable emergent states for an FRW universe with both the positive and negative spatial curvatures.
This represents that the conformal Weyl gravity can be free of the initial
 singularity problem.
Then, following Barrow {\it et al}, we address the possible types of the  finite-time future cosmological singularities.
We  discuss how these singularities also can  be avoided in the context of this theory. 
\end{abstract}
\section{Introduction}
The standard model of cosmology based on the Einstein general theory of relativity
(GR) has a good agreement with the recent high resolution observations and
provides an accurate overview of the cosmic history of our universe. However, it suffers from some fundamental problems such
the initial big bang singularity problem, the flatness problem, the horizon problem, the magnetic
monopoles and etc. Although the inflation scenario provides solutions for
some of these problems,   the initial big bang singularity problem still remains unsolved.
Indeed, if the inflation scenario is realized by the dynamics of scalar  fields coupled to Einstein gravity, then the Hawking-Penrose singularity theorem \cite{pen, vil} proves that an inflationary universe is geodesically incomplete in its past. Therefore, there  exists a singularity before
the onset of inflation and consequently the inflationary scenario
 is not capable to yield the complete history of the very early universe. 
To deal with this shortcoming, a number of attempts have been made in construction
of cosmological models which are initially non-singular. There are two main proposals
for this aim named as  the ``bouncing universe'' and ``emergent universe'' scenarios. The bouncing universe  scenario is based on a smooth transition of the universe at a finite radius from
a contracting phase  to an expanding phase. Then,  by going backward in time,
the universe collapses up to a finite value for the scale
factor, before which it starts to expand again \cite{novello}. 
Although the bouncing scenario avoids the big bang singularity, the
existence of the non-singular bounce requires non-standard matter fields violating energy conditions \cite{viol}.
In the emergent universe scenario, the universe emerges from a static state,
namely as  the ``Einstein
static universe (ESU)'', characterized by a non-zero spatial volume with a positive curvature \cite{emerg}.  This scenario avoids the initial big bang singularity while preserving the standard energy
conditions. However, due to  the existence of varieties of perturbations, such as the quantum fluctuations and the variations of the average energy density of the system \cite{ed}, the emergent universe model is unstable and suffers from a fine-tuning problem \cite{st}.
This unstability problem can be amended by the modifications
of the cosmological field equations of Einstein's GR. Indeed, there are two
conditions which any gravitational theory should satisfy for admitting an
emergent universe paradigm: $i)$ the  existence and stability
of the ESU, and $ii)$ the possibility of joining the standard cosmological history by a graceful exit mechanism from the ESU. By this motivation, the emergent universe scenario and the stability issue
of its Einstein static state has been explored in the context of various modified theories of gravity such
as  the loop quantum gravity \cite{loop}, $f(R)$ gravity \cite{fr}, $f(T)$ gravity \cite{f(T)},  Einstein-Cartan theory \cite{Cartan},   Braneworlds
\cite{brane} and Massive gravity \cite{massive} as well as in the presence
of vacuum energy corresponding to  conformally invariant fields \cite{reza} among the others.
Most of the mentioned references just focused on the first condition without
addressing the later one.

 On the other hand, after the discovery of the accelerating expansion of
the universe, deeper studies of the features of the cosmic fluid responsible for this accelerating expansion, the so called ``dark energy'', showed the plethora of new types of singularities different than the initial big bang singularity. These new types of singularities are  future singularities for a universe and they are characterized by the violation of some of the
energy conditions. This violation in turn results in the divergencies in
some  of the physical quantities such as the
scale factor, energy density and pressure profiles. As instances for these type of singularities, one may refer to  Big Rip singularity \cite{br},   Sudden singularity \cite{s} and Big Freeze singularity \cite{bf}. 
There are also other classifications for the possible future singularities,
see for example \cite{os}.

The theory of conformal Weyl gravity has been proposed as an alternative  theory to Einstein's GR
for the high energy limit and is based on the assumption of the local conformal invariance in the geometry of
spacetime \cite{ad}. There are many attempts for obtaining the Einstein's
GR as the low energy limit of the conformal Weyl theory, by dynamically breaking the conformal invariance, with varying degrees of success and difficulty. In this direction, the main problem is that by breaking the conformal symmetry, the quantum fluctuations bring back the Einstein-Hilbert action with a repulsive rather than attractive gravity. However, one may refer to the recent work of Maldacena \cite{Juan}  where it is shown that  in the  four dimension, the conformal gravity with a Neumann boundary
condition is classically equivalent to the Einstein gravity with a
cosmological constant. 
It has been  also suggested by Kazanas and Mannheim that the conformal Weyl gravity can be considered an independent  theory on its
own right instead of looking for a low energy GR limit. By this motivation,
 the exact vacuum solutions
as well as  the cosmological solutions of this theory have been found
 \cite{mk1, mk2, m3, m4}.
It is shown that in the cosmological setup, the  conformal Weyl gravity can solve naturally both the cosmological constant and the
flatness problems in GR.
Beside these successes of the conformal Weyl gravity, it is
argued that this theory fails to fulfill simultaneously the observational
constraints on the present cosmological parameters and on the primordial light element abundances \cite{yep}.  For the static setup of the theory, it is shown that the de Sitter space can be found as the vacuum solution of this theory and there is
no need to an {\it ad hoc} introduction of the cosmological constant to the gravitational action and its corresponding field equations \cite{mk1}. Moreover, in the Newtonian limit of the theory, there are  also linear modification terms to the exterior Schwarzschild-de Sitter solutions  that could explain the rotation curves of galaxies without  any need to the mysterious  dark matter \cite{mk1, man, eder}.
Another argument about this theory is given in \cite{flan}. It was  argued that the conformal Weyl gravity does not agree with the predictions
of general relativity in the limit of weak fields and non-relativistic
velocities.  Nevertheless, it was
counter-argued in \cite{mani}, that in the presence of macroscopic long range
scalar fields, the standard Schwarzschild phenomenology of GR 
is still recovered. In order to  check the viability of the conformal Weyl gravity, the deflection of light and time delay in the exterior of
a static spherically symmetric source were investigated in \cite{via}, and
it was found that the parameters of the theory fit the experimental
constraints. Also, it is shown in \cite{lobo} that one can find a class of wormhole geometries satisfying the energy conditions in the neighborhood of the throat of  wormhole. This  is in a clear contrast to the solutions in GR where exotic matter fields are needed to support a wormhole geometry.

In this work,  regarding the above mentioned interesting features of the conformal Weyl gravity, we are motivated to study the issue of the 
past and future cosmological singularities in the context
of this theory. Such a study can be also well motivated even one does not
consider the Weyl gravity theory as an independent  theory on its
own right, but as the high energy limit alternative to GR where the being
of singularities are important and inevitable issues.
  The organization of  this paper is as follows. In Section 2, we review the conformal Weyl gravity theory and its
cosmological field equations. In Section 3, we explore the existence and stability conditions for an ESU within the emergent universe scenario. In Section 4, we address the possible
future finite-time cosmological singularities.
Finally, in Section 5, we give our concluding remarks.

 \section{Conformal Weyl Gravity Theory} 
The action of the conformal Weyl gravity  is given by
\begin{eqnarray}\label{action}
I=-\alpha\int d^4x \sqrt{-g} ~C_{\mu\nu\rho\lambda}C^{\mu\nu \rho\lambda}+ \int d^4 x \sqrt{-g}~\mathcal{L}_{M},
\end{eqnarray}
where $C_{\mu\nu\rho\lambda}$ is the Weyl tensor and $\mathcal{L}_M$ is the
Lagrangian of the matter fields. Here, $\alpha$
is a dimensionless gravitational coupling constant. The conformal Weyl tensor and matter Lagrangian read respectively
as 
\begin{eqnarray}
&&C_{\mu\nu\alpha\beta}=R_{\mu\nu\alpha\beta}-g_{\mu[\alpha}R_{\beta]\mu}+\frac{1}{3}Rg_{\mu[\alpha}g_{\beta]\nu},\\
&&\mathcal{L}_M=-\int d^4 x \sqrt{-g}\left[\frac{1}{2}S^\mu S_\mu +\lambda
S^4 -\frac{1}{12}S^2 R^\mu_\mu +i\bar\psi \gamma^\mu(x)\left( \partial_\mu
+ \Gamma_\mu(x) \right)\psi -hS\bar\psi \psi    \right],
\end{eqnarray}
where $S(x)$ is a conformal scalar field that yields symmetry breaking and makes the particles to be massive, and  $\psi(x)$ is a fermion field representing
the normal matter fields. Also, $\Gamma(x), ~R^\mu_\mu,~~h $ and $\lambda$ are the fermion spin connection, scalar curvature and two dimensionless coupling
constants, respectively.
The gravitational  field equations of this theory can be obtained  by the variation of the action (\ref{action}) with respect to the metric $g_{\mu\nu}$ as \cite{mk1}
\begin{equation}\label{fe}
4\alpha W_{\mu\nu}=T_{\mu\nu},
\end{equation}
where $T_{\mu\nu}$ is the  energy-momentum tensor corresponding to the Lagrangian
of the matter fields and $W_{\mu\nu}$ is given by
\begin{equation}
W_{\mu\nu} = W_{\mu\nu}^{(2)} - \frac{1}{3}W_{\mu\nu}^{(1)} ,
\label{fieldeq}
\end{equation}
where $W_{\mu\nu}^{(1)}$ and $W_{\mu\nu}^{(2)}$ are 
\begin{eqnarray}
W_{\mu\nu}^{(1)} &=& 2g_{\mu\nu}\nabla_{\beta}\nabla^{\beta}
R^{\alpha}_{\; \alpha} -
2\nabla_{\nu}\nabla_{\mu}R^{\alpha}_{\; \alpha}
- 2R^{\alpha}_{\; \alpha} R_{\mu\nu} +
 \frac{1}{2}g_{\mu\nu}(R^{\alpha}_{\; \alpha})^2,
\nonumber \\
W_{\mu\nu}^{(2)} &=& \frac{1}{2}g_{\mu\nu}\nabla_{\beta}\nabla^{\beta}R^{\alpha}_{\; \alpha} +\nabla_{\beta}\nabla^{\beta}R_{\mu\nu} -\nabla_{\beta}\nabla_{\nu}R_{\mu}^{\; \beta} -\nabla_{\beta}\nabla_{\mu}R_{\nu}^{\; \beta} -2R_{\mu\beta}R_{\nu}^{\; \beta} +\frac{1}{2}g_{\mu\nu}R_{\alpha\beta}R^{\alpha\beta}.
\end{eqnarray}
Due to the conformal invariance of the theory both of the $T_{\mu\nu}$ and $W_{\mu\nu}$ are kinematically traceless. 

Considering the FRW metric
\begin{equation}
ds^2=-dt^2 + a^2(t)\left(\frac{dr^2}{1-kr^2} +r^2 d\Omega^2 \right),
\end{equation}
which is a conformally flat metric, one finds the Friedmann equations \cite{m3}
\begin{eqnarray}
&&\rho(t) +\lambda S^4+\frac{S^2}{2}\left( \frac{\dot a^2(t)}{a^2(t)}+\frac{k}{a^2(t)} \right)=0,\label{f1}\\
&&\rho(t)-3p(t)+4\lambda S^4 +S^2\left( \frac{\ddot a(t)}{a(t)}+ \frac{\dot a^2(t)}{a^2(t)}+\frac{k}{a^2(t)}\right)=0,\label{f2}
\end{eqnarray}
where $\rho(t)$ and $p(t)$ are the density and pressure profiles of the perfect
fluid $T_{\mu\nu}$ and $S(x)=S$  is a spacetime independent constant due
to the conformal invariance of the theory.
The covariant conservation of $T_{\mu\nu}$ leads to \cite{m3}
\begin{equation}\label{f3}
\dot \rho(t)+3H(\rho+p)=0,
\end{equation}
where $H=\dot a(t)/a(t)$ is the Hubble parameter.
 
In the next section, using the cosmological field equations (\ref{f1}), (\ref{f2}) and (\ref{f3}), we study the problem of initial big bang singularity which exists in the standard model of cosmology based on Einstein's GR.  We show that the conformal Weyl gravity admits the emergent universe scenario and then it can be free of the initial singularity problem.
\section{Emergent Universe Paradigm: The Initially Non-singular State and its Stability Analysis}
As mentioned in the introduction, the emergent universe paradigm  in GR
is unstable against prevailing perturbations in very early universe and is expected to decay rapidly \cite{st}. Therefore this scenario was developed
in modified gravity theories with the hope to improve the stability
 at the high energy regimes. Thus, the conformal
 Weyl gravity as a successful theory in solving some of the  fundamental problems in standard model of cosmology also deserves  for studying the singularity
problems in detail, which is the aim of this and the next sections. 

\begin{figure}
\centering
\includegraphics[scale=0.5]{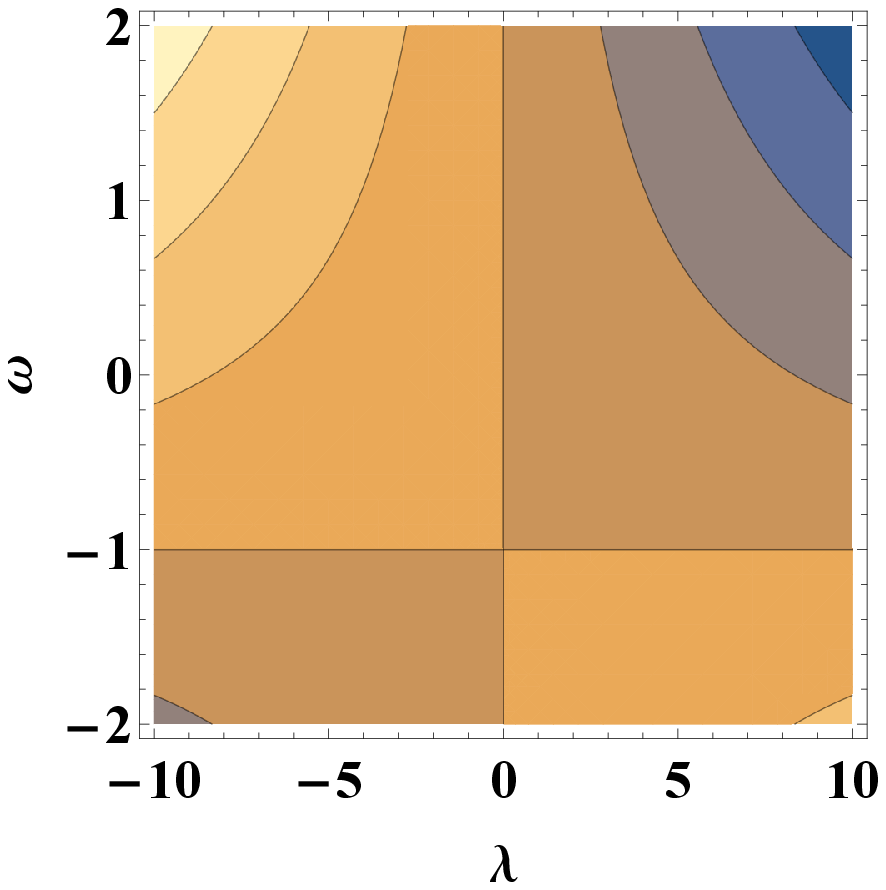}
\includegraphics[scale=0.4]{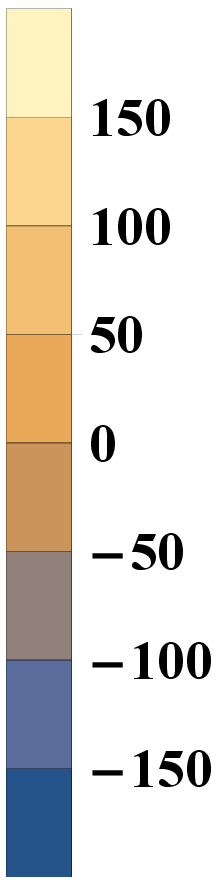}
\caption{\label{sigma2}The 2-dimensional contour plot for the eigenvalues $\sigma^2(\lambda, \omega)$ of the Jacobian
matrix $J_{ij}$. The being of $\sigma^2<0$ regions
represent the stability of the solutions.}
\end{figure}

To study the emergent universe paradigm and its stability in the context
of the conformal Weyl gravity,  here we consider  that the perfect fluid has the barotropic equation of state
$p=\omega \rho$. Then, we can combine the Friedmann equations (\ref{f1}) and (\ref{f2}) as the Raychadhuri equation
\begin{equation}\label{ray}
\ddot a(t)=-\frac{1}{2}(1+3\omega)\frac{\dot a^2(t)+k}{a(t)}-3(\omega+1)\lambda S^2a(t).
\end{equation}
The ESU is characterized by the condition $\ddot a(t)=0=\dot a(t)$ \cite{st}. Then, to begin with,
one has to obtain
the existence condition for an ESU solution in the conformal Weyl theory. The corresponding matter density
and scale factor
for ESU can be obtained from (\ref{f1}) and (\ref{ray}) as  
\begin{eqnarray}
\rho_{ES}&=&-\lambda S^4-\frac{kS^2 }{2a_{ES}^2},\label{aes}\\
a_{ES}^2&=&-\frac{(1+3\omega)k}{6(1+\omega)\lambda S^2}\label{gogo}.
\end{eqnarray}
Using (\ref{gogo}), the existence condition for an ESU solution reduces to the reality of the corresponding scale factor $a_{ES}$. Then, one can obtain the existence conditions as  they are classified  in Table \ref{existence}.
\begin{center}
\begin{table}[!ht]
\centering
\begin{tabular}{|c|c|c|} 
\hline
$\omega$ values  & $\lambda$ values &$k$ values
\\ 
\hline 
& $\lambda>0$ &$k<0$\\
\raisebox{1ex}{$\omega>-1/3$}
& $\lambda<0$ &$k>0$\\
\hline
& $\lambda>0$ &$k>0$\\
\raisebox{1ex}{$-1<\omega<-1/3$}
& $\lambda<0$ &$k<0$\\
\hline
& $\lambda>0$ &$k<0$\\
\raisebox{1ex}{$\omega<-1$}
& $\lambda<0$ &$k>0$\\
\hline
\end{tabular}
\caption{\label{existence}The existence conditions for an ESU in the context
of conformal Weyl gravity.}
\end{table}
\end{center}
Then, the following points can be realized regarding (\ref{aes}), (\ref{gogo}) and Table \ref{existence}.
\begin{itemize}
\item One notes  that since the scale factor of the
initial ESU $a_{ES}$ has very small size, the corresponding density in (\ref{aes}) will be extremely high unless the scalar field $S$  is very small too. In turn, this represents that the matter density $\rho_{ES}$ can be
arbitrary small too as the consequence of the smallness of $S$. This is one of the unique and  interesting properties  of the fourth order
conformal Weyl gravity. The interpretation is that  the Universe can born from nothing and any cosmological inhomogeneities such as those due to the large scale  structures represent a perturbation around the vacuum state $T_{\mu\nu}=0$, which   has mentioned  also by Mannheim \cite{m3}.
 Indeed, this is a direct result of the vanishing rank four Weyl tensor $C_{\mu\nu\alpha\beta}$ and consequently the rank two gravitational
tensor $W_{\mu\nu}$ in the field equation of conformal Weyl gravity  (\ref{fe}) for an FRW metric. There is no such an interesting
 property for an FRW universe in the context of the second order Einstein's GR theory. 
\item In the context of Weyl gravity there exists the possibility of having
ESU for the ordinary matter fields possessing the equation of state  $\omega\geq 0$, and there is no need for exotic matter fields.
\item The case of $\omega=-1$, representing the cosmological constant field, does not correspond to  an small size initial emergent ESU for the conformal
Weyl gravity theory.
Indeed, as the equation of state parameter $\omega$ approaches to $-1$, we have $a_{ES}\rightarrow \infty$. This fact has two meanings: $i)$ there is an extremely large static de Sitter limit for the theory,  and $ii)$ by approaching the equation of state parameter $\omega$ to $-1$, possibly due to the perturbation in $T_{\mu\nu}$ in very early universe \cite{m3}, the initial small size non-singular Einstein static state  enlarges and enters to an inflationary phase which makes  hint to a natural  graceful exit mechanism. At the end of this section, we will discuss  this case again.
\item There is no non-singular  spatially flat ESU for the conformal Weyl theory, i.e., for $k=0$. \item The case of $\omega>-1/3,~\lambda>0$ and $k<0$ has a particular importance in the sense that it can provide  a solution for the flatness problem in the standard model of cosmology, see \cite{m3}.
\item The case $\omega=-1/3$  represents a singular state, i.e., $a_{ES}=0$.
Then, any perturbation around $T_{\mu\nu}=0$ toward the  equation of
state $\omega=-1/3$ for matter fields yields
an initially singular universe for the conformal Weyl gravity.
\end{itemize}
Now, to study the stability of ESU, we consider the Raychadhuri equation (\ref{ray})
as the following  two dimensional dynamical system  with the phase-space variables  
$x_1=a(t)$ and $x_2=\dot a(t)$
\begin{eqnarray}\label{dyn}
&&\dot x_1(t)=x_2(t),\nonumber\\
&&\dot x_2(t)=-\frac{1}{2}(1+3\omega)\frac{x_2^2(t)+k}{x_1(t)}-3(\omega+1)\lambda S^2x_1(t).
\end{eqnarray}
Then, for this dynamical system, ESU is corresponding to the fixed point
$(x_1, x_2)=\left(a_{ES}, 0\right)$. The stability of this equilibrium point
can be realized by determining eigenvalues $\sigma^2$ of the Jacobian
matrix $J_{ij}=\frac{\partial \dot x_i}{\partial x_j}|_{x_2=0}$.
The case $\sigma^2>0$ represents a hyperbolic fixed point, and hence it will
be unstable. For this case, the trajectories in the neighborhood of this
point diverge from it. In contrast, $\sigma^2<0$  represents a stable center equilibrium point in which a small departure form this point results in oscillations
about it. Indeed this case corresponds to the initial stable Einstein static state in which
the seed universe oscillates around
it indefinitely. The  Jacobian matrix associated to our dynamical system
(\ref{dyn}) reads as
\begin{equation}
J_{ij}=\begin{pmatrix}0 & 1 \\
\frac{(1+3\omega)k}{2a_{ES}^2}-3(1+\omega)\lambda S^2 & 0 \\
\end{pmatrix},
\end{equation}
where its eigenvalues can be obtained as
\begin{equation}\label{sigma}
\sigma^2=-6(\omega+1)\lambda S^2,
\end{equation}
where we have used (\ref{gogo}) in (\ref{sigma}).  In Figure \ref{sigma2}, we give the two dimensional contour  plot representing $\sigma^2(\lambda, \omega)$. The presence of $\sigma^2(\lambda,\omega)<0$
regions imply the stability of the solutions.
Then, one finds that there are possible stable solutions satisfying the following
stability conditions 
\begin{equation}\label{stab}
\text{ Stability conditions:}\left\{\begin{array}{ll}
         \displaystyle \omega<-1
         &\mbox{for $\lambda<0$,}\\
         &\\
         \displaystyle \omega>-1
         &\mbox{for $\lambda>0$}
\end{array} \right.
\end{equation}
Then, regarding the conditions in Table \ref{existence} and in (\ref{stab}), one can conclude the required
conditions guaranteing both the  existence and
stability of an ESU in the conformal Weyl gravity as are given in Table \ref{exs}.
\begin{center}
\begin{table}[!ht]
\centering
\begin{tabular}{|c|c|c|} 
\hline
$\omega$ values & $\lambda$ values &$k$ values
\\ [.5ex]
\hline 
$\omega>-1/3$& $\lambda>0$ &$k<0$\\
\hline
$-1<\omega<-1/3$& $\lambda>0$ &$k>0$\\
\hline
$\omega<-1$& $\lambda<0$ &$k>0$\\
\hline
\end{tabular}
\caption{\label{exs}The conditions guaranteing both of the existence and stability of an ESU in the conformal Weyl gravity theory.}
\end{table}
\end{center}

\begin{figure}
\centering
\includegraphics[scale=0.6]{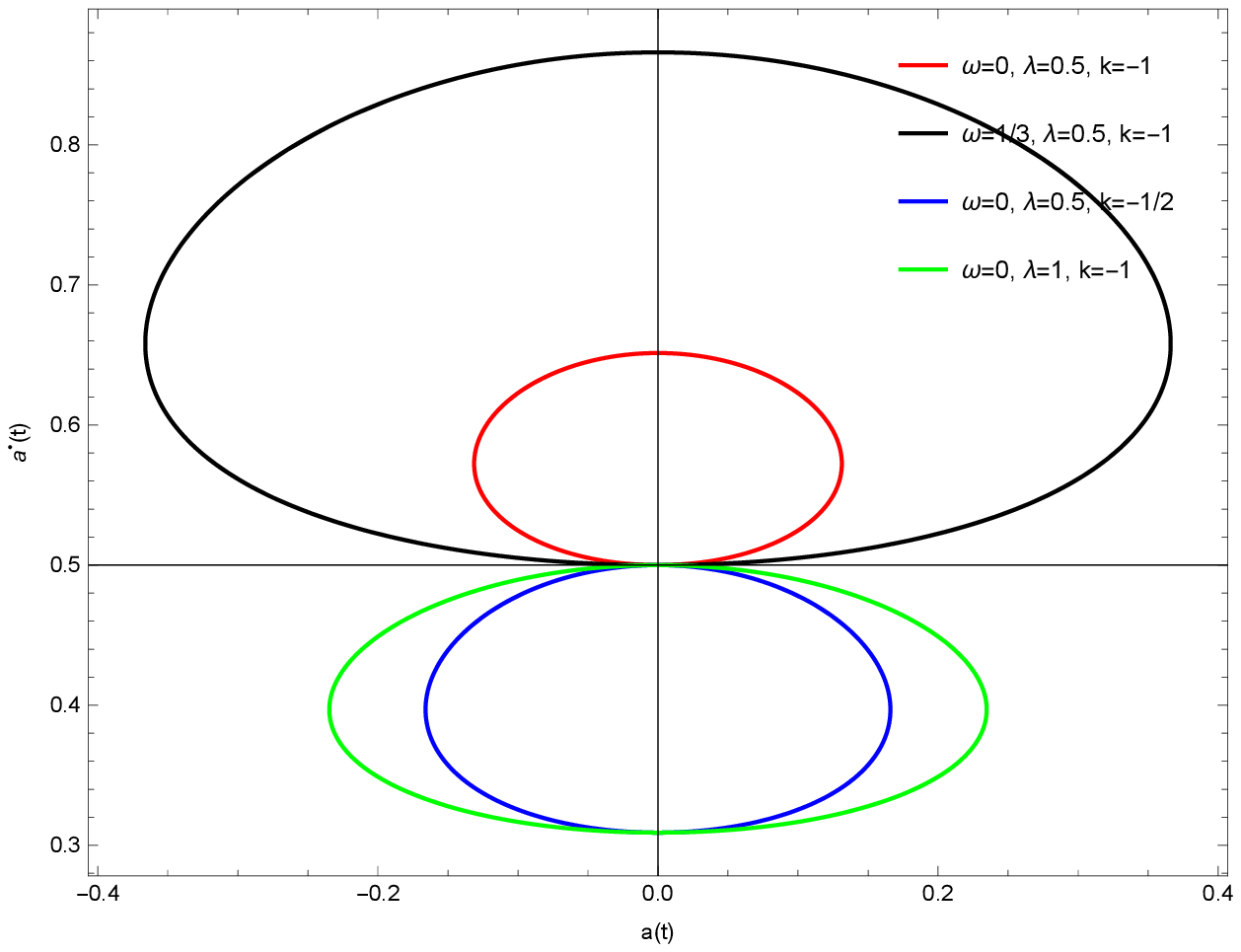}
\includegraphics[scale=0.7]{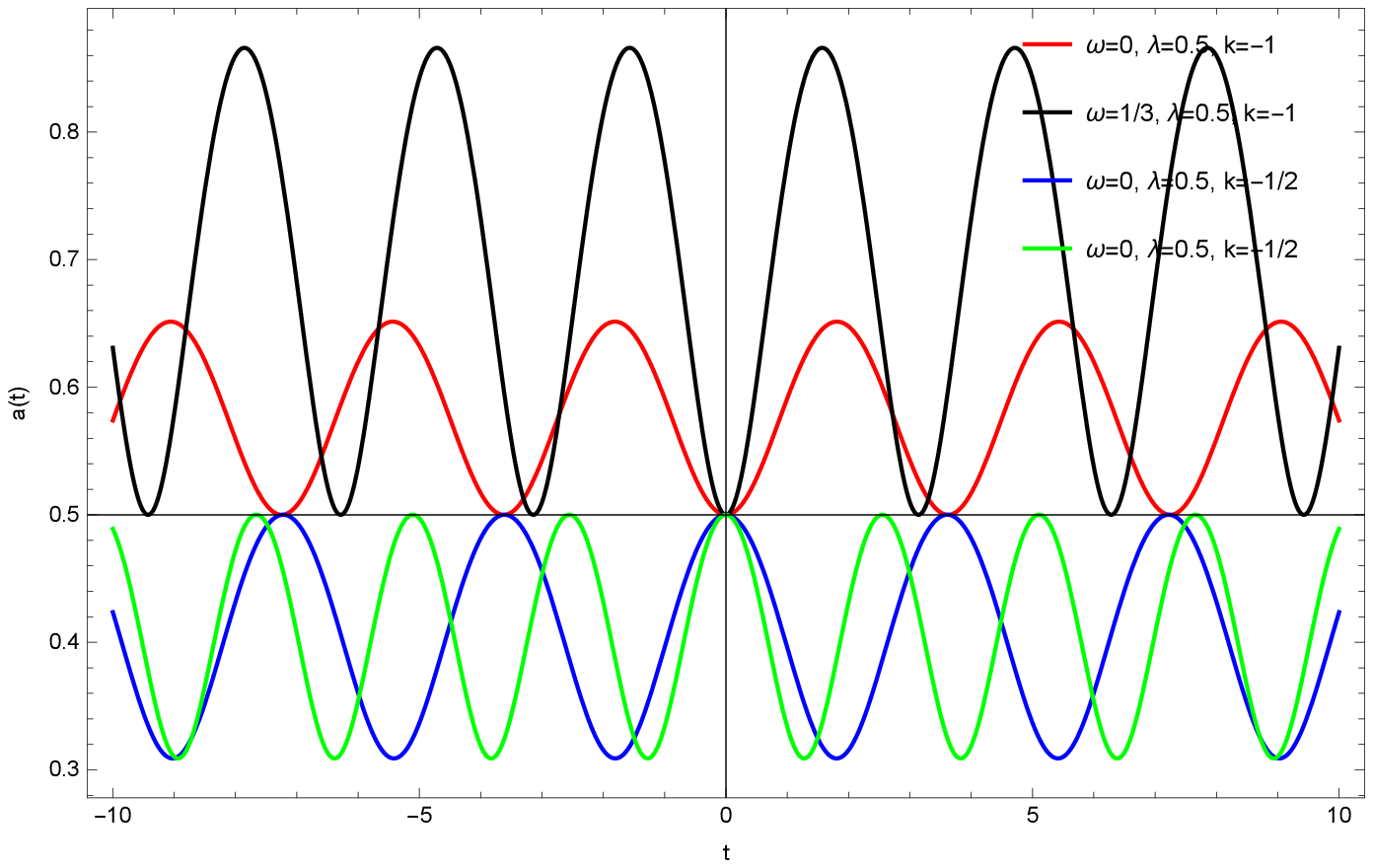}
\caption{\label{os}The upper plot represents some closed trajectories in the phase-space $\left(\dot a(t),  a(t)\right)$ for the given typical values of the parameters satisfying the existence and stability conditions in Table \ref{exs}.   The lower plot represents the evolution of the corresponding scale factors (as a function of time) to each of the closed trajectories in the phase space. The corresponding cases are in the same color.}
\end{figure}

In Figure \ref{os}, we have plotted the phase-space $(\dot a(t), a(t))$ diagram and the evolution of the scale factor for the corresponding closed trajectories in the phase-space.  The upper plot represents some closed trajectories in the phase-space for the given typical values of the parameters satisfying the existence and stability conditions in Table \ref{exs}. In general, a closed trajectory  in the phase space, as a limit cycle, represents
an oscillatory system \footnote{The differential equation (\ref{dyn}) possesses
the Jacobi elliptic function $x_1=A\,cn(\omega t+ \delta)$ as its solution
which satisfies the equations $x_1^\prime=(1-x_1^2)(1-k_1^2 +k_1^2x_1^2)$ and $x_1^{\prime\prime}=-(1-2k_1^2)x_1-2k_1^2x_1^3$ where $k_1$ is an arbitrary
constant. In the case of $k_1=0$, the solution reduces to the form $x_1=A\cos(\omega
t +\delta)$.}. The lower plot represents the evolution of the corresponding oscillatory scale  factors, versus the time, to each of the closed trajectories in the phase-space. The corresponding cases are denoted by the same colors
and the same set of parameters.

Until here, we have proved that the conformal Weyl gravity admits an stable
ESU for the emergent universe scenario. As we addressed before, the matter
density supporting this ESU can be arbitrary small, representing the possibility for being
born from nothing \cite{m3}. Beside the possibility of being born from nothing for the universe,  the ESU scenario in the conformal Weyl gravity has one another important theoretical advantage in comparison to the ESU in
Einstein gravity as well as to some of the other modified gravity theories. This advantage is related to the fact that   the de Sitter solution is a
vacuum solution of the conformal Weyl theory. This vacuum solution is capable of triggering the inflation such that under the prevailing perturbations the universe
can transit from the oscillations around its small size ESU to its subsequent inflationary phase and then produces whole of the cosmic history 
\cite{mk1}. This advantage  is  supported by the fact that for both the FRW
solution and  the de Sitter solution (represented by $R_{\mu\nu}=\Lambda g_{\mu\nu}$),  both of the Weyl tensor $C_{\mu\nu\alpha\beta}$
and the gravitational tensor $W_{\mu\nu}$ vanish. Then, as mentioned by
Mannheim \cite{m3}, the FRW cosmology is also the case that the energy momentum tensor is everywhere zero, and then it is another vacuum solution of the theory. Thus, any cosmological inhomogeneities such as those in the large-scale structure of the universe is a consequence of a perturbation around $T_{\mu\nu}=0$. In the presence of normal matter, such
a property  can be achieved nontrivially by an interplay between the
various fields of the theory.  In  the conformal Weyl gravity, the perturbations around $T_{\mu\nu}=0$
are responsible for both the generation the matter fields ($\rho(t)$ and $p(t)$) associated to the inhomogeneities in the large scale structures and the transition from one non-singular vacuum state to its subsequent another vacuum state deriving the inflation.  This unique feature of having these two vacuum solutions in the conformal Weyl
gravity may provides  a natural mechanism for the graceful exit from the ESU to its subsequent inflationary phase and then to the radiation and matter
dominated eras.  We aim to study and elaborate on this important advantage of the conformal Weyl theory and report  it elsewhere \cite{my}. 

Before concluding this section, we  consider an specific solution: a non-singular
solution for  $\omega=1/3$ and $\rho(t)=\rho_0/a^4(t)$ which represents a radiation filled universe.
For this case, the Friedmann equation (\ref{f1}) can be integrated which gives the following solution for the  scale factor
\begin{equation}\label{rad}
a(t)=\sqrt{-\frac{k}{4\lambda S^2}-\sqrt{\frac{k^2 -16\rho_0\lambda}{16\lambda^2 S^4}}\,\cos\left(\sqrt{8\lambda S^2}\,t \right)}.
\end{equation}
This solution represents the oscillation of the  scale factor   of the radiation universe between two finite minimum
and maximum sizes. Then, one can
interpret this solution in two different ways as
\begin{description}
\item[$i)$] An ever oscillating radiation dominated universe model with the finite minimum and maximum sizes.
\item[$ii)$] An eternally oscillating small size emergent universe  standing as the seed of a large scale universe. In this case,
 the large scale universe can born and exit from this finite size  state
and then experience an inflationary phase by the
prevailing perturbations around $T_{\mu\nu}=0$. 
\end{description}

Believing in the history of the universe having subsequent phase transitions from the radiation
and dust to the dark energy dominated phases seems to be more in the favor of the above second interpretation. This is also
supported by the current observations about the accelerating expansion of
the universe in the sense that the repulsive dark energy profile  makes unlikely the being of a contracting phase.
  Here,  regarding the square roots in (\ref{rad}), one sees that to have the oscillatory modes  around  the scale factor 
$a_0=\sqrt{-\frac{k}{4\lambda S^2}}$, the conditions $k<0$ and $\lambda>0$ are required. These conditions 
are in agreement with our results for an emergent ESU given in Table \ref{exs}. Also it is seen that
$\rho$ should satisfy the condition $\rho_0<k^2/16\lambda$ which represents
that the initial density of  ESU state is finite for a finite and non-zero values of $k$ and
$\lambda$ parameters, respectively.
\section{ The Possible Future Singularities}
In this section, following Barrow \textit{et al} \cite{Barrow1, Barrow2}, we study the possible types of finite-time singularities may happen in the future of an FRW universe in the context of the conformal
Weyl gravity.  To do this, based on our study in the past section for the
possibility of having an initially non-singular cosmology in the context
of this theory, we construct initially non-singular scale factors
and then investigate the behavior of $H(t),~\ddot a(t),~\rho(t)$  and $p(t)$ versus the cosmic time. 

In Table \ref{typesing}, we give some types
of possible future singularities for an FRW universe regarding the behavior
of the scale factor, matter density and pressure in a finite cosmological time \cite{nojir}.
\begin{center}
\begin{table}[!ht]
\centering
\begin{tabular}{|c|c|c|c|c|} 
\hline
Singularity Type & $t$ &$a(t)$& $\rho(t)$& $p(t)$
\\ [.5ex]
\hline 
I& $t\to t_s$ &$a\to \infty$& $\rho\to \infty$ &$\mid p\mid \to \infty$\\
\hline
II&  $t\to t_s$ &$a\to a_s$& $\rho\to \rho_s$& $\mid p\mid  \to \infty$\\
\hline
III&  $t\to t_s$ &$a\to a_s$& $\rho\to \infty$ &$\mid p\mid \to \infty$\\
\hline
IV&  $t\to t_s$ &$a\to a_s$& $\rho\to 0$ &$\mid p\mid \to 0$\\
\hline
\end{tabular}
\caption{\label{typesing}The possible finite time future singularities for an FRW cosmology.}
\end{table}
\end{center}
Types I, II, III and IV singularities are known as ``Big Rip or Cosmic Doomsday''\cite{cald}, ``Sudden'' \cite{Barrow1}, ``Big Freeze'' \cite{freeze} and ``Big Brake or
Big Separation'' \cite{sep} singularities,
respectively. In type IV singularity, although $a(t),~\rho(t)$ and $p(t)$ remain finite but the higher derivatives of $H$ diverge.  Following \cite{Barrow1, Barrow2},  and to keep the generality
of the study in this section, we do not impose any specific
equation of state on the pressure and density profiles in the field equations
(\ref{f1}) and
(\ref{f2}). We consider the solution for the scale factor
$a(t)$ introduced by Barrow \cite{Barrow1, Barrow2} 
\begin{equation}
a(t)=1+B t^q +C (t_s -t)^n,
\end{equation}
where  $B,~q,~C$ and $n$ are positive free constants to be determined, and
$t_s$ represents the occurring moment of the singularity.
For an initially singular universe, i.e possessing the big bang singularity,  one can
fix the zero of time by the condition $a(0)=0$ which leads to $C=-1/t_s^n$
and  the following form for the scale factor
\begin{equation}
a(t)=1+(a_{s} -1) \left(\frac{t}{t_s}\right)^q -\left( 1-\frac{t}{t_s}\right)^n.
\end{equation}
where $q>0$ and $n>0$. In the present work, regarding the previous section
showing the possibility of having initially non-singular
universe in the context of the conformal Weyl gravity, we consider the following
generalization for the above scale factor
\begin{equation}\label{a}
a(t)=1+(a_{s} -1) \left(\frac{t}{t_s}\right)^q +\left(a_{ES}-1\right)\left( 1-\frac{t}{t_s}\right)^n,
\end{equation}
where we supposed $0<a_{ES}\ll 1<a_{s}$ and it  describes a non-singular universe evolving from  $a(0)=a_{ES}$, where $a_{ES}$ represents the
initially non-singular Einstein static state, toward  $a(t_s)=a_s$.  One can simply recover
the initially singular universe scenario by setting  $a_{ES}=0$. 
Then, using (\ref{a}), we obtain 
\begin{eqnarray}
\dot a(t)&=&\frac{q(a_s -1)}{t_s}\left( \frac{t}{t_s}\right)^{q-1} -\frac{n(a_{ES}-1)}{t_s}\left( 1-\frac{t}{t_s}\right)^{n-1}, \\
\ddot a(t)&=&\frac{q(q-1)(a_s -1)}{t_s^2}\left( \frac{t}{t_s}\right)^{q-2} +\frac{n(n-1)(a_{ES}-1)}{t_s^2}\left( 1-\frac{t}{t_s}\right)^{n-2}.
\end{eqnarray}
On the other hand, using the field equations (\ref{f1}) and (\ref{f2}),
we find the density and pressure as
\begin{eqnarray}
&&\rho(t) =-\lambda S^4-\frac{S^2}{2}\left( \frac{\dot a^2(t)}{a^2(t)}+\frac{k}{a^2(t)} \right),\label{f1*}\\
&&p(t)=\lambda S^4 +\frac{1}{3}S^2 \frac{\ddot a(t)}{a(t)}+ \frac{1}{6}S^2
\left(\frac{\dot a^{2}(t)}{a^2(t)}
+\frac{k}{a^2(t)}\right),\label{f2*}
\end{eqnarray}
Here, one notes to the difference
between the field equations of the fourth-order conformal Weyl gravity
and the second-order Einstein gravity. The sign differences in $\rho(t)$ and $p(t)$ is a natural result
of the difference between the actions of these two theories. Using this difference
in (\ref{f1*}), it was shown by Mannheim \cite{m3} that the conformal Weyl gravity is free of the flatness problem.  Another interesting feature of the conformal
Weyl gravity is the coupling way of the scalar field $S$ to the dynamical quantities
$a(t)$ and $\dot a(t)$ which can play an important role in the future singularity
avoidance as we will discuss later.  

Considering that the coupling $\lambda$ and the scalar field $S$ are supposed
to be constant in the conformal Weyl
gravity \cite{m3}, we can classify the possible finite time future singularities in the context of
this theory with respect to $n$ and $q$
values as in the following cases.
\begin{itemize}
\item \textbf{The case of $t \to t_s$ with $0<n<1$ and $0<q\leq1$}.
\\
For this case we have
\begin{eqnarray}
&&a(t_s)\to a_s,~~\dot a(t_s)\to +\infty,~~ H(t_s)\to  +\infty,~~\ddot a(t_s)\to +\infty,\\&&\rho(t_s)  \to -\infty,~~ |p(t_s)|\to \infty.
\end{eqnarray}
This case represents the type III singularity as in the table (\ref{typesing}),
or  Big Freeze singularity. For this case, the evolution of the scale factor $a(t)$, density $\rho(t)$ and pressure $p(t)$ are plotted in Figure \ref{a-rho-p1} for two typical values of $q$. 
\begin{figure}
\centering
\includegraphics[scale=0.58]{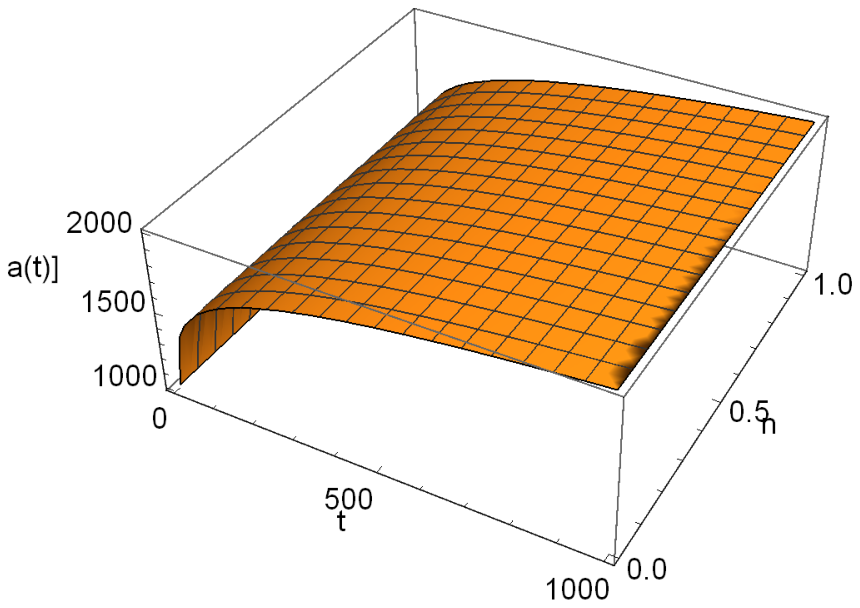}
\includegraphics[scale=0.58]{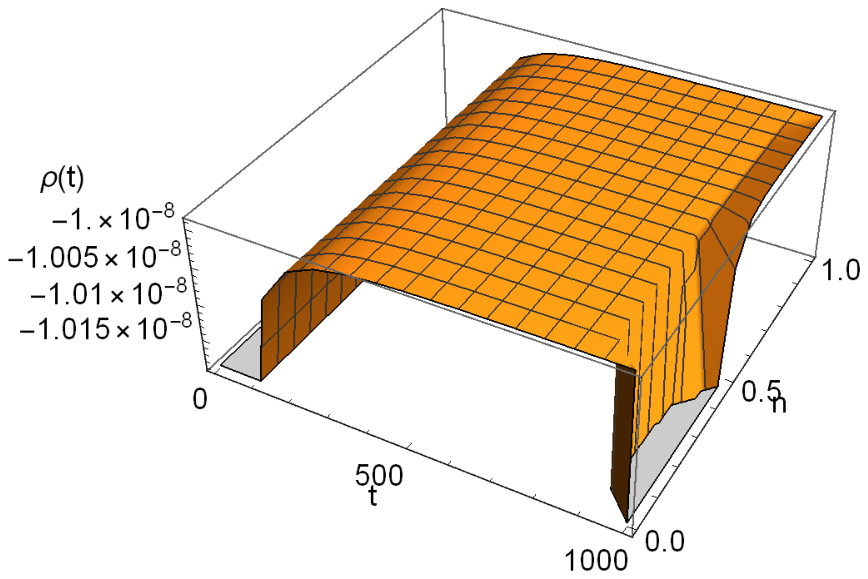}
\includegraphics[scale=0.58]{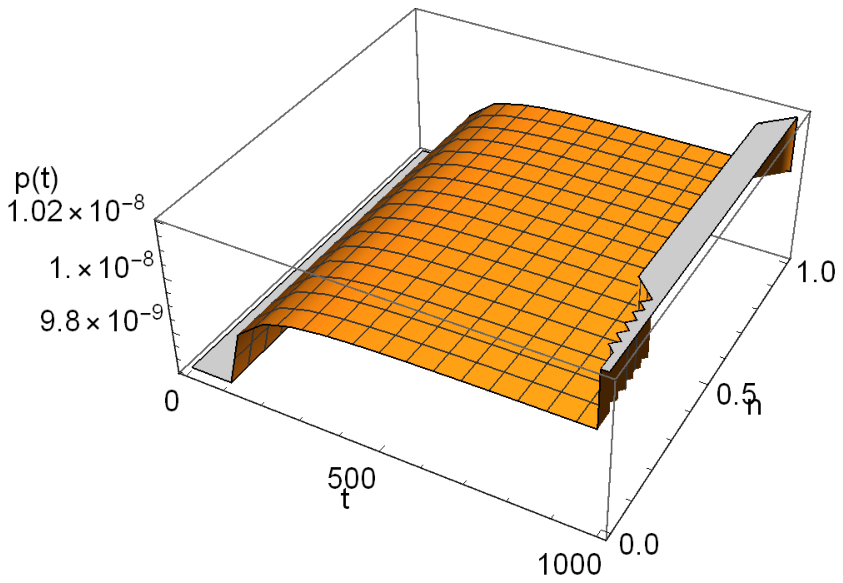}
\includegraphics[scale=0.58]{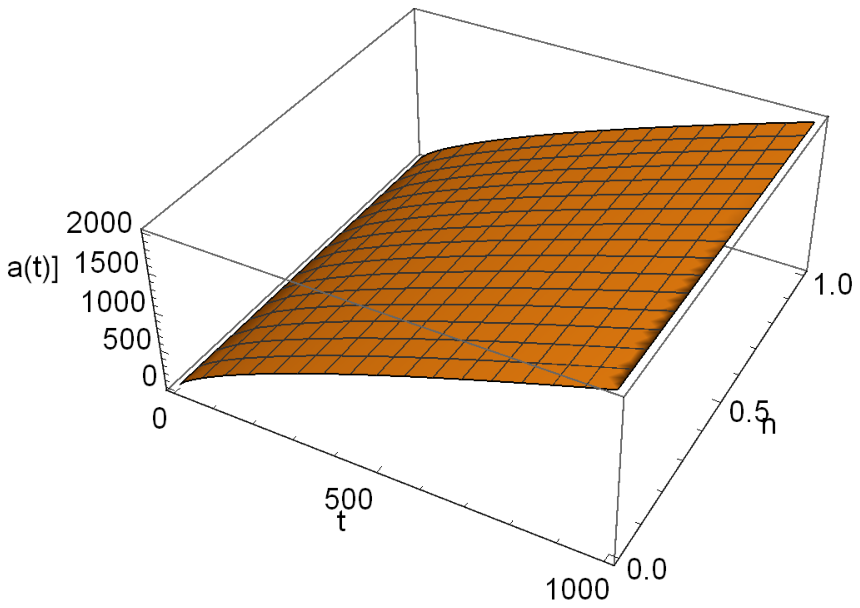}
\includegraphics[scale=0.58]{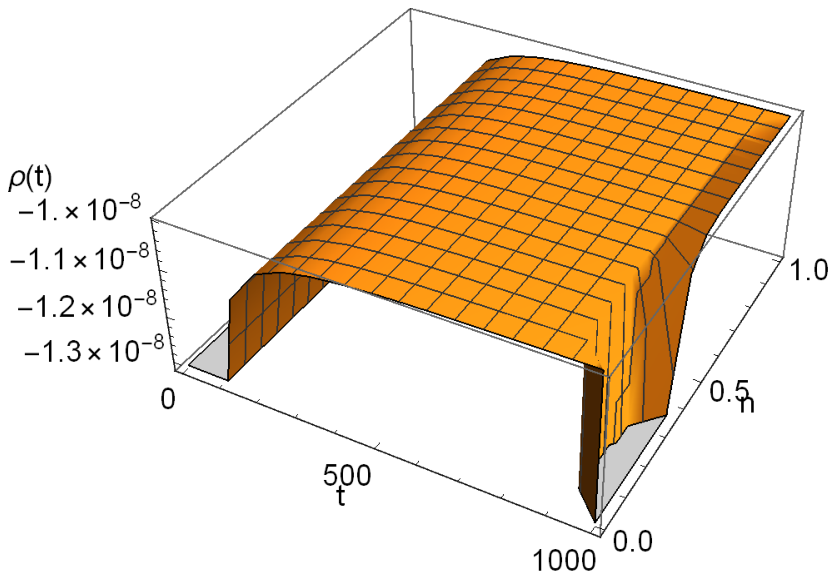}
\includegraphics[scale=0.58]{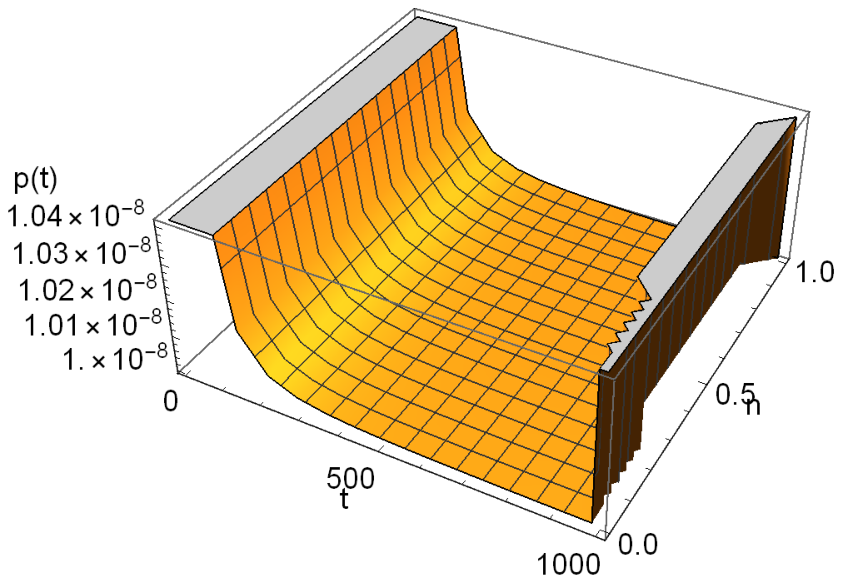}
\caption{\label{a-rho-p1}The evolution of the scale factor $a(t)$, density $\rho(t)$ and pressure $p(t)$ for typical values of $q=0.2$ in the upper plot and $q=0.8$ in the lower plot. We have set $a_s=2000,~~
a_{ES}= 0.01,~~t_s= 1000,~~\lambda= 1,~~S = 0.01$ and 
$k = -1$. We see that by increasing $q$ value, the slopes of the $\rho$ and $p$ plots for $t\to t_s$ is increasing which means $\rho$ and $p$ blowing up rapidly. The initial apparent over-density and pressure in the
lower plots are due to the smallness of $a_{ES}$.}
\end{figure}
\item \textbf{The case of $t \to t_s$ with $n=1$ and $0<q\le1$.}
\\
For this case we find
\begin{eqnarray}
&&a(t_s)\to a_s, ~~\dot a(t_s)\to \dot a_s>0,~~ H(t_s)\to H_s>0,~~\ddot a(t_s)\to \ddot a_s\leq0\\
&&\rho(t_s)  \to \rho_s<0,~~ |p(t_s)|\to p_s
\end{eqnarray}
where all the $a_s,~\dot a_s,~H_s,~\ddot a_s,~\rho_s$ and $p_s$ are finite.
Then, there is no finite-time future singularity for this case.
Here, the zero acceleration case happens for $q=1$. Similarly, the evolution of the scale factor $a(t)$, density $\rho(t)$ and pressure $p(t)$ are plotted in Figure \ref{a-rho-p2} for two typical values of $q$.
\begin{figure}
\centering
\includegraphics[scale=0.54]{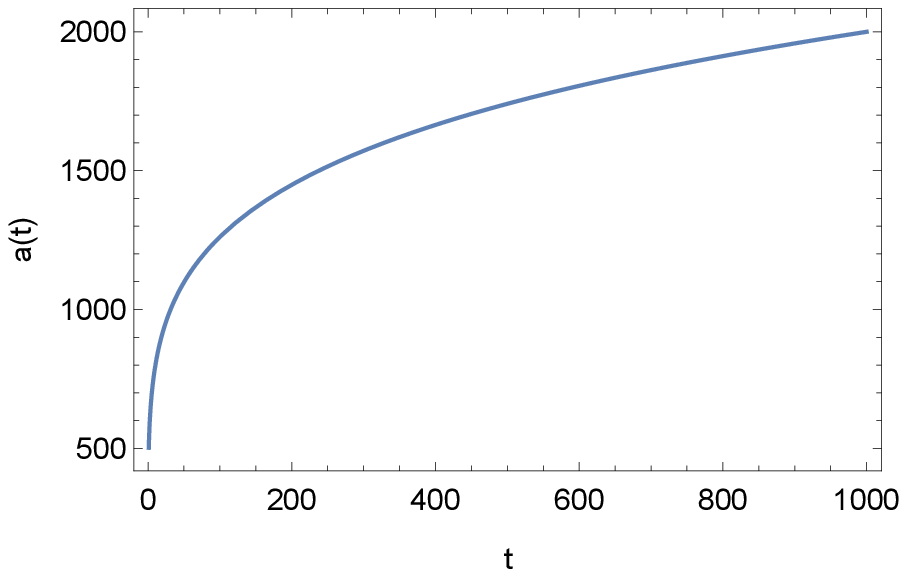}
\includegraphics[scale=0.64]{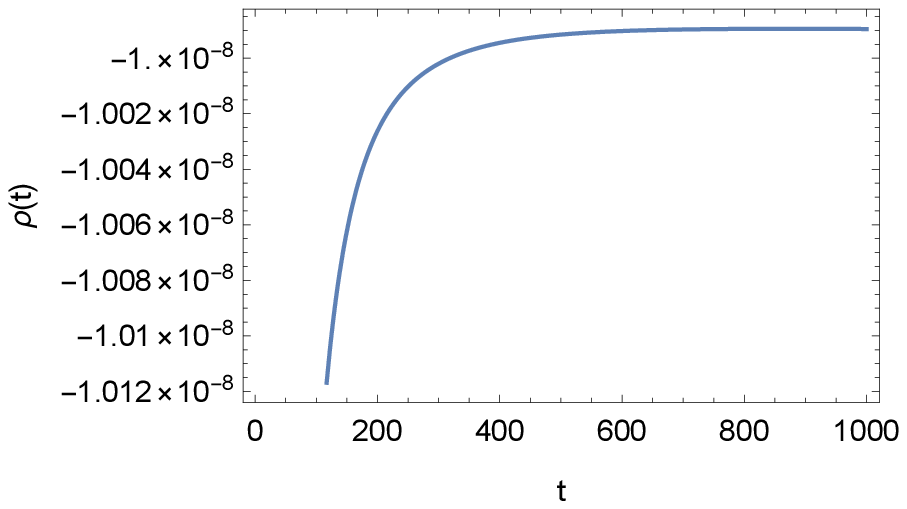}
\includegraphics[scale=0.6]{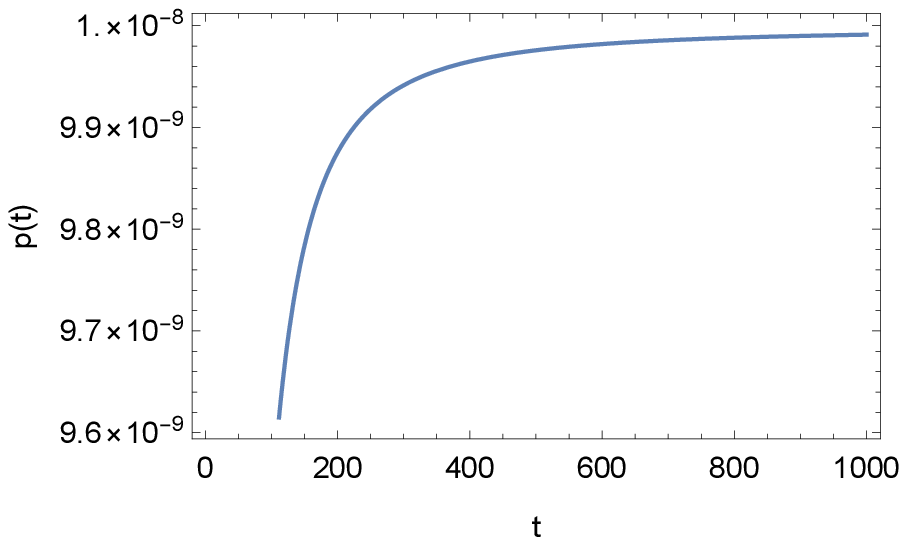}
\includegraphics[scale=0.54]{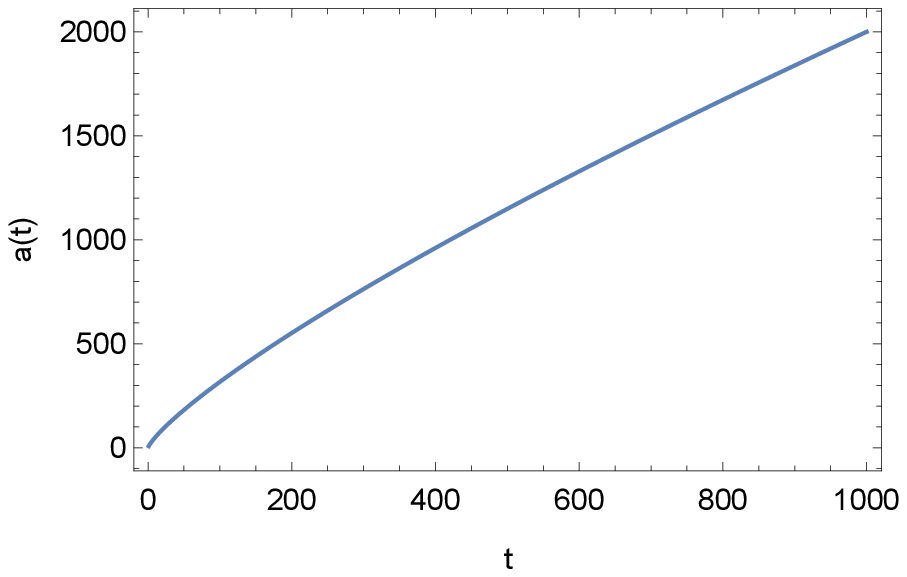}
\includegraphics[scale=0.62]{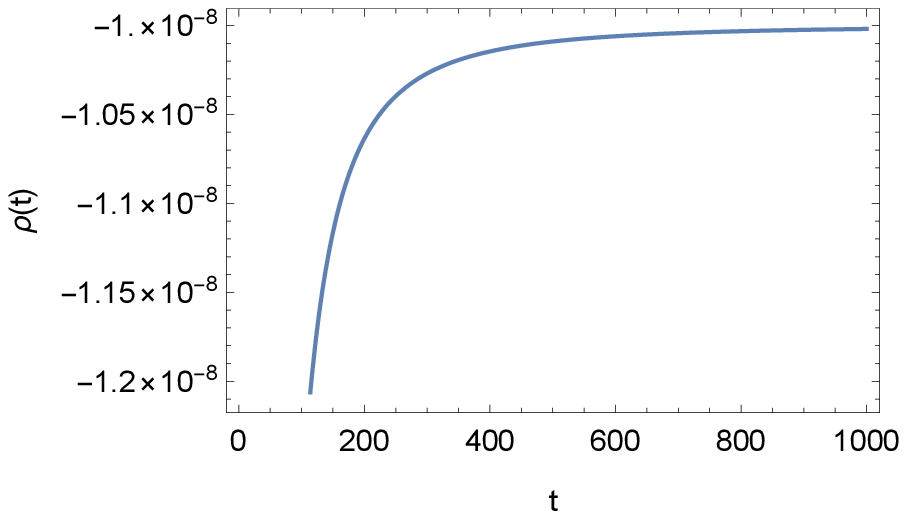}
\includegraphics[scale=0.6]{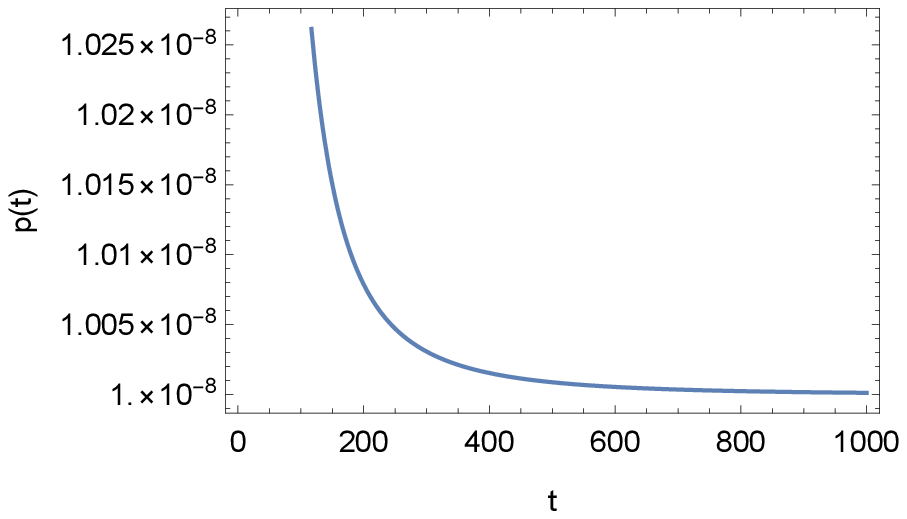}
\caption{\label{a-rho-p2}The evolution of the scale factor $a(t)$, density $\rho(t)$ and pressure $p(t)$ for typical values of $q=0.1$ in the upper plot and $q=0.8$ in the lower plot.  We have set $a_s=2000,~~
a_{ES}= 0.01,~~t_s= 1000,~~\lambda= 1,~~S = 0.01$ and 
$k = -1$.  It is seen that by increasing $q$ value, the slopes of the $\rho$ and $p$ plots
for $t\to t_s$ is increasing. The initial over-density and pressure  are due to the smallness of $a_{ES}$.}
\end{figure}
\item \textbf{The case of $t \to t_s$ with $1<n<2$ and $0<q\leq1$.}
\\
For this case we have
\begin{eqnarray}
&&a(t_s)\to a_s,~~\dot a(t_s)\to \dot a_s>0,~~ H(t_s)\to  H_s>0,~~ \ddot a(t_s)\to -\infty\\
&&\rho(t_s)  \to \rho_s <0,~~ |p(t_s)|\to \infty.
\end{eqnarray}
This case represents a type II singularity, i.e. a cosmological sudden singularity.
For this case, the evolution of the scale factor $a(t)$, density $\rho(t)$ and pressure $p(t)$ are plotted in Figure \ref{a-rho-p3} for typical values of $q$.
\begin{figure}
\centering
\includegraphics[scale=0.58]{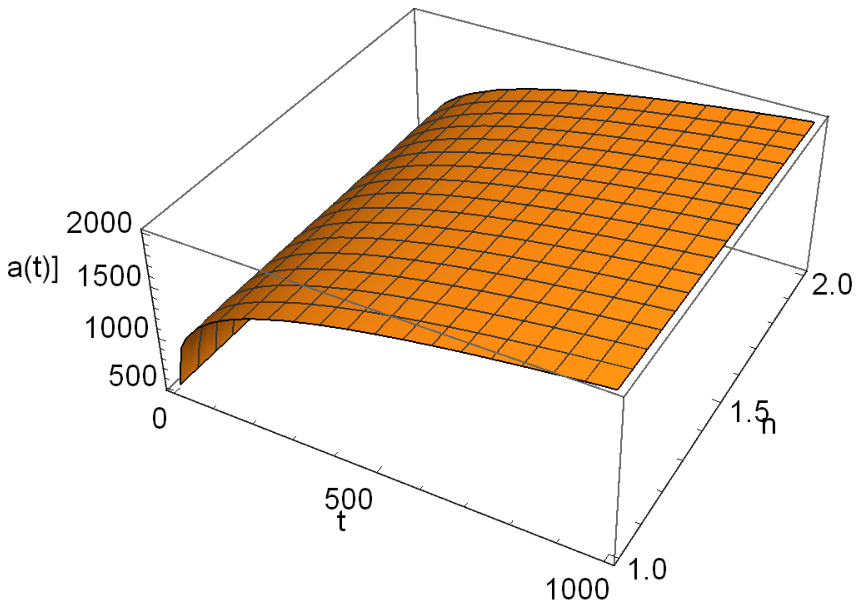}
\includegraphics[scale=0.58]{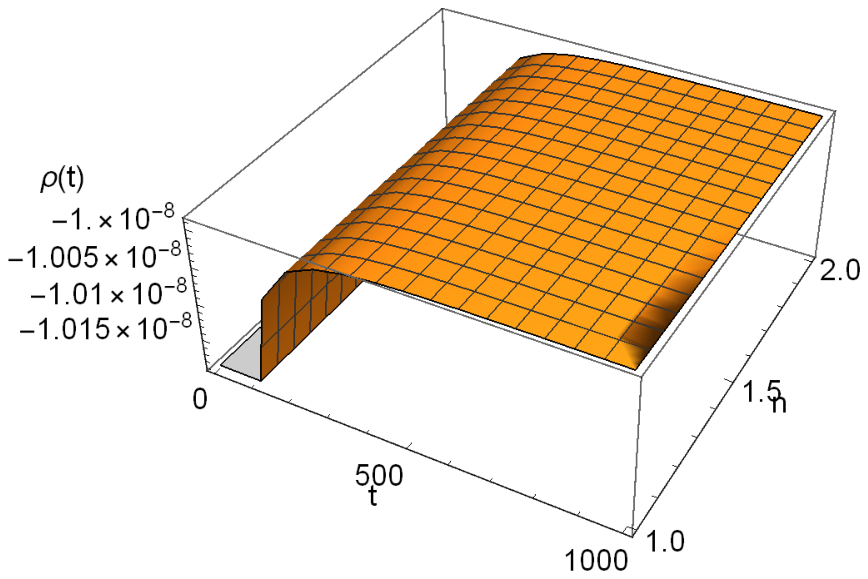}
\includegraphics[scale=0.58]{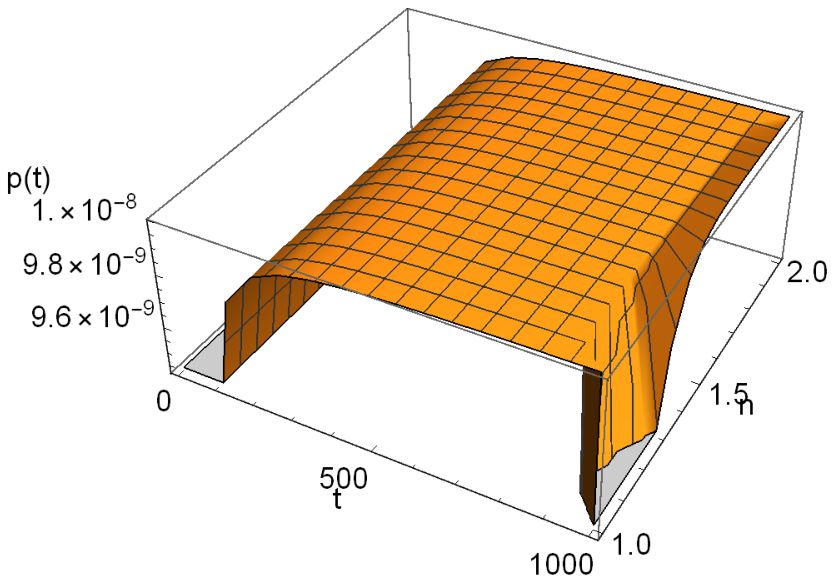}
\includegraphics[scale=0.58]{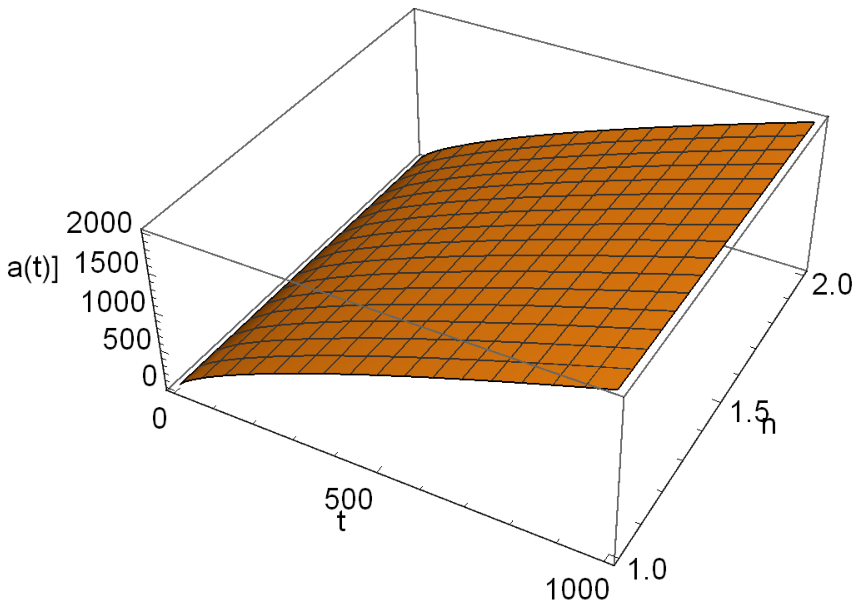}
\includegraphics[scale=0.58]{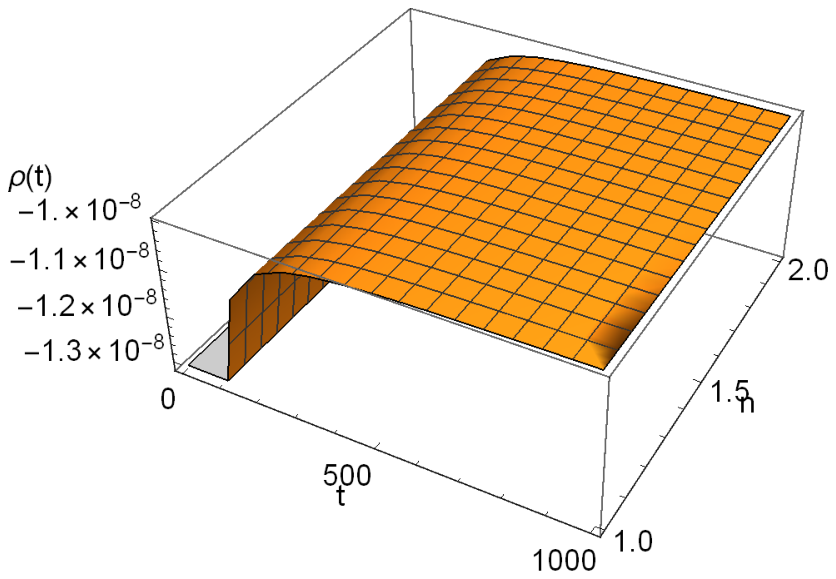}
\includegraphics[scale=0.58]{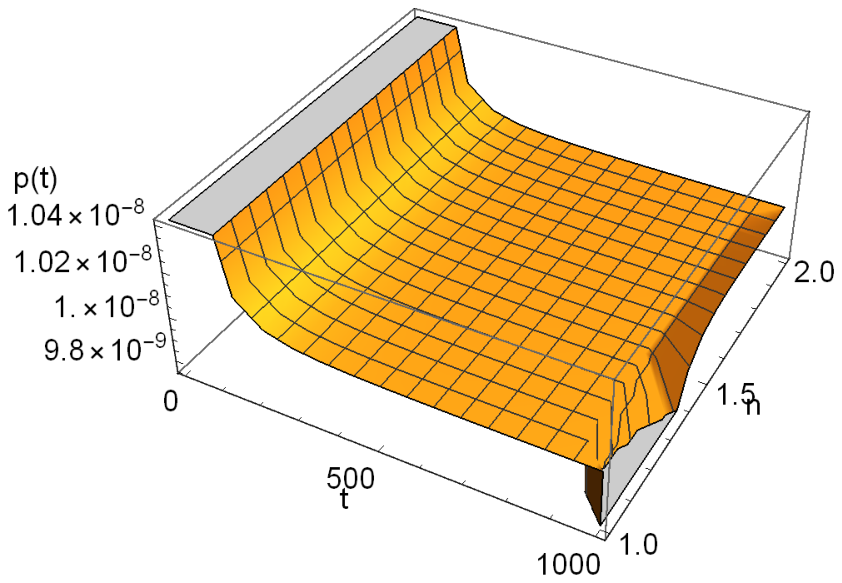}
\caption{\label{a-rho-p3}The evolution of the scale factor $a(t)$, density $\rho(t)$ and pressure $p(t)$ for typical values of $q=0.2$ in the upper plot and $q=0.8$ in the lower plot.  We have set $a_s=2000,~~
a_{ES}= 0.01,~~t_s= 1000,~~\lambda= 1,~~S = 0.01$ and 
$k = -1$.  By increasing $q$ value,  $\rho$ and $p$ increase more rapidly
for $t\to t_s$. The initial over-density and pressure in the
lower plots are due to the smallness of $a_{ES}$.}
\end{figure}
\item  \textbf{The case of $t \to t_s$ with $n\geq2$ and $0<q\leq1$.}
\\
For this case we have
\begin{eqnarray}
&&a(t_s)\to a_s,~~\dot a(t_s)\to \dot a_s>0,~~ H(t_s)\to  H_s>0,~~ \ddot a(t_s)\to \ddot a_s\leq0\\
&&\rho(t_s)  \to \rho_s<0,~~ |p(t_s)|\to p_s.
\end{eqnarray}
where all $a_s,~\dot a_s,~H_s,~\ddot a_s,~\rho_s$ and $p_s$ are finite.
Then, for this case all the physical quantities remain finite.
The evolution of the scale factor $a(t)$, density $\rho(t)$ and pressure $p(t)$ are plotted in Figure \ref{a-rho-p4} for two typical values of $q$.
\begin{figure}
\centering
\includegraphics[scale=0.5]{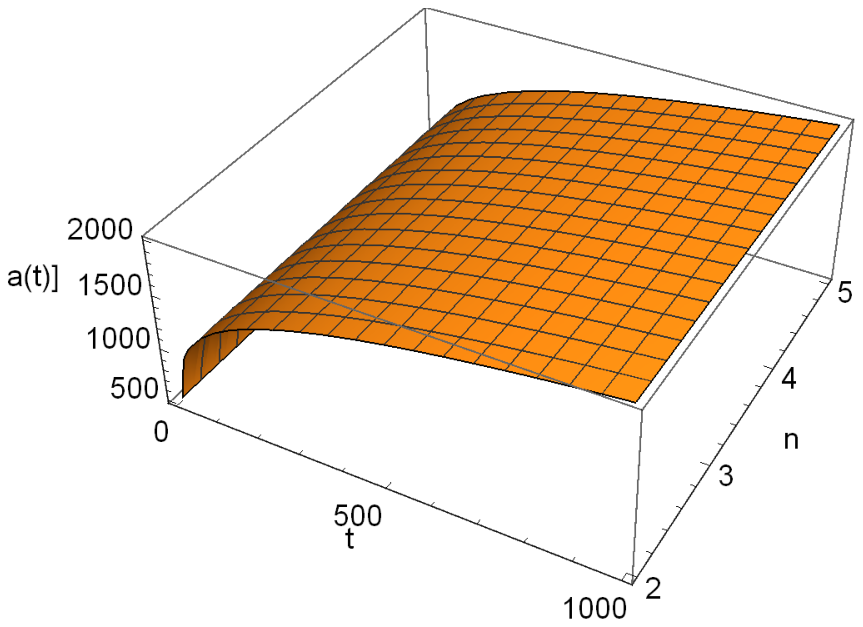}
\includegraphics[scale=0.5]{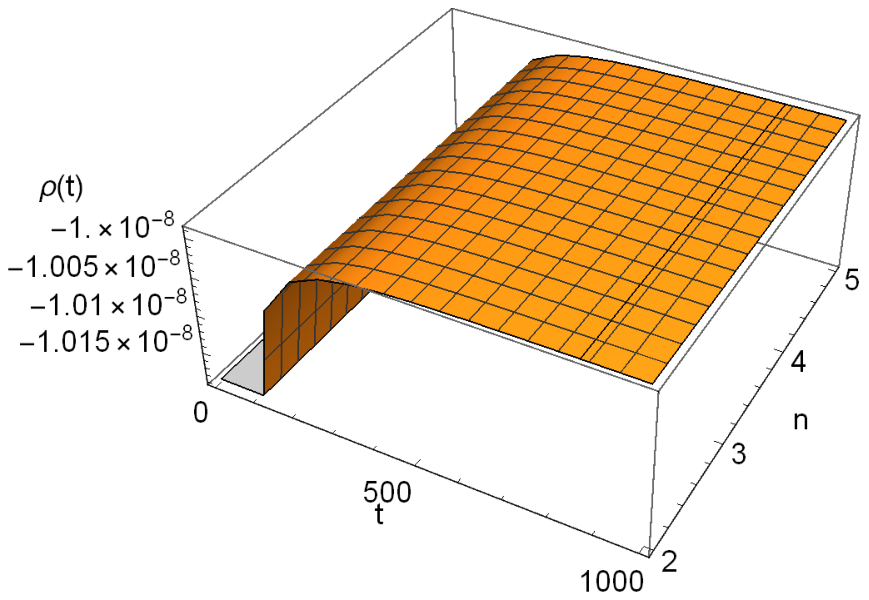}
\includegraphics[scale=0.5]{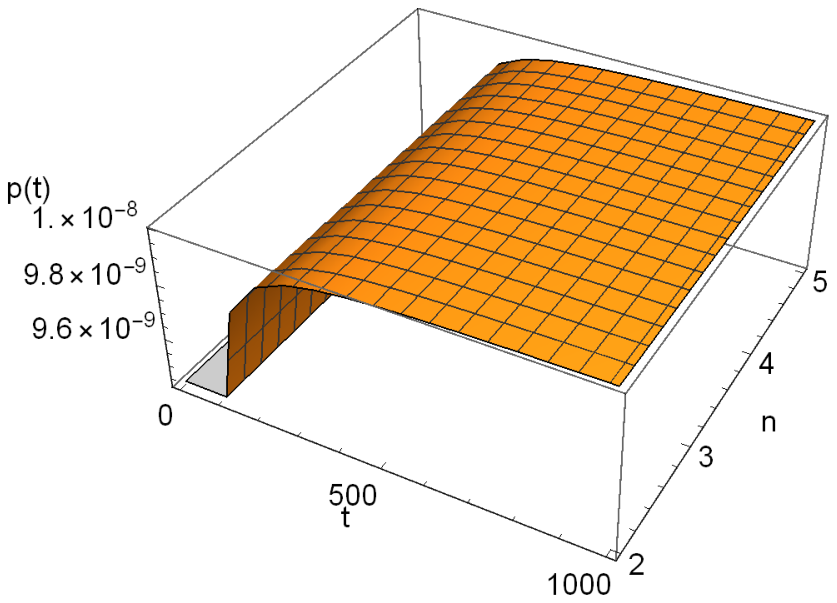}
\includegraphics[scale=0.5]{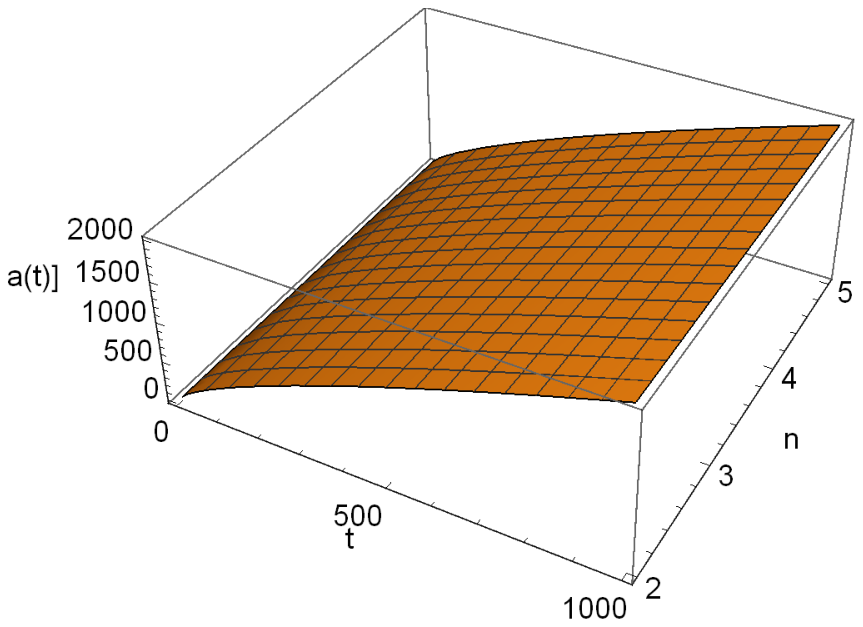}
\includegraphics[scale=0.5]{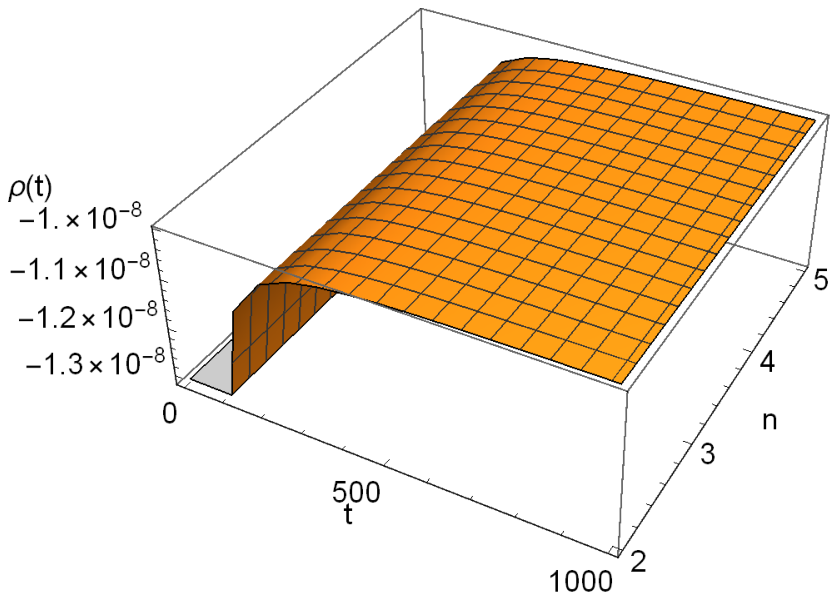}
\includegraphics[scale=0.5]{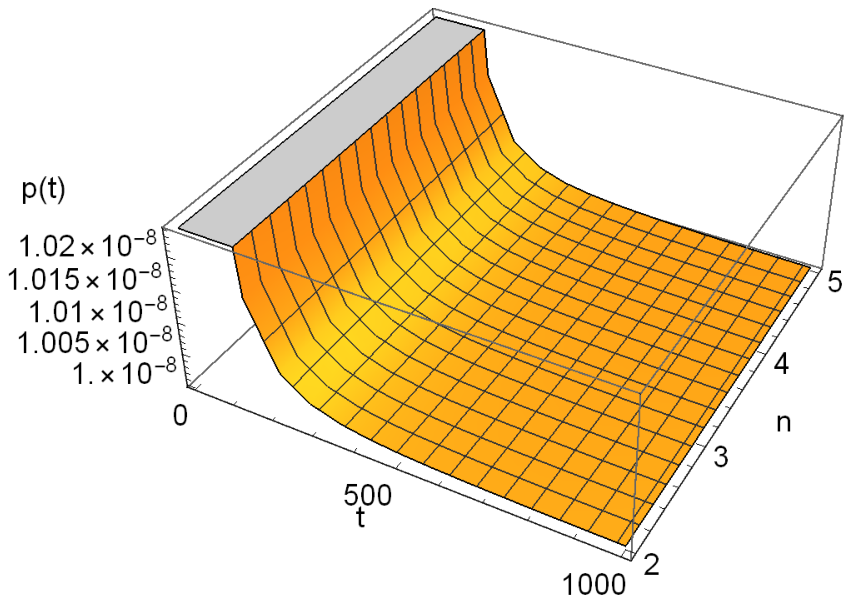}
\caption{\label{a-rho-p4}The evolution of the scale factor $a(t)$, density $\rho(t)$ and pressure $p(t)$ for typical values of $q=0.2$ in the upper plot and $q=0.8$ in the lower plot. We have set $a_s=2000,~~
a_{ES}= 0.01,~~t_s= 1000,~~\lambda= 1,~~S = 0.01$ and 
$k = -1$.  By increasing $q$ value, $\rho$ and $p$ increase rapidly
for $t\to t_s$. The initial over-density and pressure in the
lower plots are due to the smallness of $a_{ES}$.}
\end{figure}
\\
Then, we see that for an initially non-singular  universe which evolves according
to (\ref{a}), in general there are type
II and III cosmological singularities in the context of conformal Weyl gravity. The type IV singularity, i.e $t\to t_s,~a\to a_s,~\rho\to 0$ and $|p|\to 0$, in the context of this theory with the scale factor  (\ref{a}) can be achieved for the very small values of the scalar field $S$ values  ($S\to
0$) such that $S \dot a_s\to
0$ and $S^2 \ddot a_s\to 0$. Even in the cases of divergent $\dot a(t)$ and $\ddot a(t)$, the limiting values of $S \dot a_s\to
0$ and $S^2 \ddot a_s\to 0$ can be obtained as the result of  $0\times \infty $ type limits
which  require $\frac{\ddot a(t)}{\dot a^2(t)}$ and $\frac{\dddot a(t)}{\ddot a^2(t)}$  to be  finite, respectively.
Finally, for the scale factor (\ref{a}),  for $0\leq t\leq t_s$ we have $a_{ES}\leq
a(t)\leq a_s$, then the big rip singularity defined as $t\to t_s,~a\to \infty,~\rho\to \infty$ and $|p|\to \infty$ can not be achieved
for a finite time. However,  one may  consider the scale factor
\begin{equation}\label{ripi}
a(t)=a_{ES}\left(1+\frac{t}{t_s -t}  \right)^n,
\end{equation}
 for a universe evolving from an initial  non-singular state $a_{ES}$ where $0\leq t \leq t_s$ and $n\geq1$. Then,  through the field equations (\ref{f1*})
and (\ref{f2*}), for $t\to t_s$, we find $a\to \infty,~\rho\to -\infty$ and $|p|\to \infty$ representing a Big Rip type singularity.
\\
We conclude our analysis for the possible future cosmological singularities in the context
of the conformal Weyl gravity in the following two points.
\end{itemize}
\begin{description}
\item[$i)$] 
In general, there are possibilities for    having the types II,
III and I cosmological singularities  for an initially non-singular  universe  evolving according to the scale factors   (\ref{a}) and (\ref{ripi}),
respectively. The type IV singularity requires that the scalar field $S$
takes very small values and  $\frac{\ddot a(t)}{\dot a^2(t)}$ and $\frac{\dddot a(t)}{\ddot a^2(t)}$  to be  finite.
 Then, the final fate of the cosmos in the conformal Weyl gravity can be a sudden, big freeze, big rip or big brake singularity depending on the evolution
of the scale factor $a(t)$ and scalar field $S$ value.
 \item[$ii)$] The conformal Weyl gravity possesses one interesting property: \textit{the unique type of coupling the scalar field $S$ to the dynamical quantities $a(t),~H(t)$,  and $\ddot a(t)$}.  Indeed, these dynamical quantities  are
the origins of various type of  future cosmological singularities. For the singularities that the divergencies
in $\rho(t)$ and $p(t)$ are due to the divergencies in $H$ and $\ddot a(t)$,
one may consider the following finite-time fine-tunings: $S\,H(t_s)\to finite$ and $S\,\ddot a(t_s)\to finite$. Then, by these fine-tunings,  the scalar field $S$ can control and regularize the corresponding singularities and
makes free the conformal theory from any singularity problem.  The smallness
of $S$ as a result of being the FRW solution as the vacuum solution for the conformal Weyl theory guarantees the avoidance of finite time future singularities.  There is no such a possibility in Einstein's GR for a scalar field as the matter source of the field equations. Then,
the conformal Weyl gravity can provide a cosmological model which is free
of both the past and future singularities.
\end{description}
\section{Conclusion}
In the present work, we studied the issue of the past and future cosmological singularities
in the context of the fourth-order conformal Weyl gravity. For the past singularity
problem, we investigated the emergent universe scenario proposed by
Ellis {\it et al}. We obtained the stability conditions for the corresponding ESU using the fixed point
approach. We showed that depending on the values of the parameters of the conformal Weyl gravity theory, there are possibilities for having initially stable ESUs for an FRW universe with both the positive and negative spatial curvatures. In particular, it is found that there exists the possibility of having ESU for the ordinary matter fields possessing the equation of state  $\omega\geq 0$, but the conformal Weyl  theory is not capable to have a non-singular  spatially flat ESU. In contrast to GR, here we found that the matter density of the initial  ESU, i.e  $\rho_{ES}$, can be
arbitrary small which represents the possibility of being born form nothing. Moreover, the fact that both the FRW
solution and  de Sitter solution are vacuum solutions of the conformal
Weyl theory provides a unique feature for this theory in the sense that the
initially non-singular emergent universe can gracefully exit from one vacuum to the other vacuum solution.
We discussed that by approaching the equation of state parameter $\omega$ to $-1$, possibly due to the perturbation in $T_{\mu\nu}=0$
in the early universe \cite{m3}, the initial small size non-singular ESU  enlarges and enters to an inflationary phase.
 In  the conformal Weyl
theory, the perturbations around $T_{\mu\nu}=0$ are responsible for both the generation of the matter fields ($\rho(t)$ and
$p(t)$) associated to the inhomogeneities in the large scale structures and for the transition from one non-singular
vacuum state to its subsequent another vacuum state deriving the inflation. This unique feature of having
these two vacuum solutions in the conformal Weyl gravity may provides a natural mechanism for a graceful
exit from the ESU to inflationary phase and then to its subsequent radiation and matter dominated eras. We aim to study and elaborate on this point and report such a possibility elsewhere \cite{my}.
 
Then, following Barrow {\it et al}, we addressed the possible types of the  finite-time future cosmological singularities such as Big Rip, Sudden, Big Freeze and Big Brake singularities.
We discussed that these future cosmological singularities can be the fate
of the universe in the context of this theory in a general setup. 
However, the conformal Weyl gravity possesses one interesting property: the
unique type of coupling of the scalar field $S$ to the dynamical quantities $a(t),~H(t)$,  and $\ddot a(t)$.  Indeed, these dynamical quantities  are
the origins of various type of  future cosmological singularities. For the singularities that the divergencies
in $\rho(t)$ and $p(t)$ are due to the divergencies in $H(t)$ and $\ddot a(t)$, by considering the  finite-time fine-tunings $S\,H(t_s)\to finite$ and $S\,\ddot a(t_s)\to finite$, the scalar field $S$ can control and regularize the corresponding singularities and
make free the conformal Weyl theory from any singularity problem.  The smallness
of $S$ as a result of being the FRW solution as the vacuum solution for the conformal Weyl theory guarantees the avoidance of finite time future singularities.  There is no such a possibility in Einstein's GR for a scalar field as the matter source of the field equations. Then,
the conformal Weyl gravity can provide a cosmological model which is free
of both the past and future singularities.

\section*{Acknowledgement}
The author deeply thanks  Prof. Metin Gürses for useful comments.

\end{document}